\documentstyle[numreferences]{klu}

\def\bfmat #1{\mbox{{\boldmath $#1$}}}

\begin{opening}
\title{Hamilton's Principle and Rankine-Hugoniot Conditions
for General Motions of Fluid Mixtures
}
\author{Henri \surname{Gouin}}
\author{Sergey \surname{Gavrilyuk}}

\institute{Universit\'e d'Aix-Marseille, Facult\'e des Sciences et Techniques,\\
Laboratoire de Mod\'elisation en M\'ecanique et Thermodynamique,
Case 322,\\
Avenue Normandie-Niemen, 13397 Marseille Cedex 20, France}
\date{E-mail: henri.gouin@univ-cezanne.fr; sergey.gavrilyuk@univ-cezanne.fr}

\end{opening}

\begin{document}

\bigskip

{\abstract{
In previous papers [1-2], we have presented hyperbolic governing equations
and jump conditions
for barotropic fluid mixtures. Now we extend our results to the most
general case
of two-fluid conservative mixtures taking into account the entropies of
components. We obtain governing
equations for each component of the medium. This is not a system of
conservation laws. Nevertheless, using Hamilton's principle
we are able to obtain a complete set of Rankine-Hugoniot conditions.
In particular, for the gas dynamics they coincide with classical jump
conditions of conservation of momentum and energy. For the two-fluid case,
the jump relations do not involve the conservation of the total momentum
and the total energy.}}

\bigskip

\noindent
{\bf Sommario.} \
In precedenti lavori [1-2] sono state dedotte equazioni di governo
iperboliche e condizioni di salto per miscele
fluide barotropiche. In questo lavoro tali risultati sono estesi al caso
pi\`u generale di miscele
di due fluidi conservative, tenendo conto delle entropie dei componenti. Si
ottengono equazioni di governo
per ciascun componente della miscela. Pur non essendo queste
un sistema di leggi di conservazione, usando il principio di
Hamilton si ottiene un insieme completo di condizioni di salto di
Rankine-Hugoniot. In particolare per la
gasdinamica, queste coincidono con le condizioni di salto
classiche per la conservazione del momento e dell'energia. Nel
caso dei due fluidi,
le condizioni di salto non coinvolgono la conservazione del momento e
dell'energia totali.

\bigskip

{\keywords{Hamilton's principle, Jump conditions, Multiphase flows.\footnote{\emph{Extended version of the paper:  "Meccanica 34: 39-47, 1999".}}}

\section{Introduction}

Hamilton's principle is a well-known method
to obtain the equations of motion in conservative
fluid mechanics (see, for example, [3-6]). It is less known that this
variational method is suitable to obtain the
Rankine-Hugoniot conditions through a surface of discontinuity [7-9].
The variations of Hamilton's action are constructed
in terms of virtual motions of continua. The virtual motions may be defined
both
in Lagrangian and Eulerian coordinates [3,10]. Such virtual motions yield
the governing
equations in different but equivalent forms. However the shock conditions
are not equivalent.
For example, for the gas dynamics, Hamilton's principle in the Lagrangian
coordinates yields the conservation of momentum and energy.
In the Eulerian coordinates, we obtain only the conservation of energy and
the conservation of
the tangential part of the velocity field [7,8]. Here we use variations
in the Lagrangian coordinates.

We assume that Hamilton's action is defined with the help of a Lagrangian
function
which is the difference between the kinetic energy and a potential
depending on
the densities, the entropies and the relative velocity
of the components of the mixture. This potential can be
interpreted as a Legendre transformation of the internal energy [1-2,11].

In [1-2,11], we have considered only the case of molecular mixtures. The
heterogeneous fluid, when each component occupies only a part of the mixture
volume, can
be described by the same system if one of the component is incompressible.
Indeed, if the phase ''1'' is
incompressible, the average density $\ \rho_1\ $ is related with the volume
concentration of
component $\ \varphi_1\ $ by $\rho_1=\rho_{10} \varphi_1, $
where $\ \rho_{10}=\ $ const is the physical density of the phase ''1''.
Hence, the
knowledge of the average densities gives the volume concentrations and the
physical
densities. In the present paper, we do not distinguish these two cases. We
shall call both cases ''two-fluid mixtures''.

It is well known (see for example Stewart and Wendroff [12]) that the
governing equations of two-fluid mixtures are not generally in a divergence
form. In this case we may not obtain the shock conditions for the system.
Moreover, the system is often
non-hyperbolic, which means the ill-posedness of the Cauchy problem. The
hyperbolic
two-fluid models were constructed by many authors (see for example [13]). The
problem to obtain the Rankine-Hugoniot conditions was an open question.
This is the aim of our paper.
By using Hamilton's principle in non-isentropic case we obtain the
governing equations
for each component and a complete set of Rankine-Hugoniot
conditions
generalizing those obtained in [1-2] for barotropic motions. To present
the basic ideas, we consider first
in section 2 the one-velocity case and extend this approach in sections 3
and 4 to the two-fluid mixtures.

Let us use asterisk "*" to denote $conjugate$ (or $transpose$) mappings
or covectors (line vectors). For any vectors \ ${\bf a},
{\bf b}$ \ we shall use the notation \ ${\bf a}^* {\bf b}$\ for their
\ $ scalar \ product $
(the line vector is multiplied by the
column vector) and \
\ ${\bf a} {\bf b}^*$\ for their
\ $ tensor \ product $
(the column vector is multiplied by the line vector).
The product of a mapping \ $A$ \ by a vector \ ${\bf a}$\ is denoted by \
$A\ {\bf a} $. Notation \ ${\bf b}^* \ A $ \ means covector
\ ${\bf c}^* $ \ defined by the rule \ ${\bf c}^* = ( A^*\ {\bf b} \ )^*$.
The divergence of a linear transformation \ $A$ \ is the covector $\ {\rm
div} A\ $ such that, for any constant vector \ ${\bf a}, $
$$
{\rm div} (A)\ {\bf a} \ = \ {\rm div }\ (A\ {\bf a} \ ).
$$
Let $\ A \ $ be any linear mapping defined on $\ \Omega_0\ $ and $\ B\ =
\ \displaystyle\ {\partial {\bf z}\over \partial {\bf Z}}\ $ be
the Jacobian matrix associated with the change of variables $\ {\bf z}\ =\
{\bf M}({\bf Z})$,
${\bf z}$ belongs to $\Omega$.
Then,
$$
\hbox{div}_0\ {A}\ =\ \det B\ \hbox{div } \pmatrix{\ \displaystyle
{B\over \det B}\ A},
\eqno{(1.1)}
$$
where ${\rm div}_0$ ( div ) means the divergence operator
in $\Omega_0$ ( $\Omega$ ). Equation (1.1) plays an important role.

The identical transformation is denoted by $I$, and the gradient line
(column) operator by $\nabla$
($\nabla^*$). For divergence and gradient operators in time-space we use
respectively symbols Div and Grad.

The elements of the matrix $A$ are denoted be $a_j^i$ where $i$ means lines
and $j$
columns. The elements of the inverse matrix $A^{-1}$ are denoted by $\bar
{a}_j^i $. If $f(A)$ is a scalar function of $A$, the
matrix $\ \ \displaystyle{{\partial f\over\partial A}}\ \ $ is
defined by the formula
$$
\displaystyle{\left({\partial f\over\partial A}\right)_i^j = {\partial
f\over \partial a_j^i}}.
$$

The repeated latin indices imply summation.
Index $\alpha = 1, 2$ refers to the parameters of components:
densities $\rho_{\alpha}$, velocities ${\bf u_{\alpha}}$, etc.

\bigskip

\section{One velocity fluid}

The consequences of this section are well known. We obtain the classical
governing equations
and the Rankine-Hugoniot conditions for the gas dynamics. Nevertheless, the
presented method is
universal and is extended for two-fluid mixtures in the following sections.

Let $\displaystyle\ {\bf z}=\pmatrix{t\cr {\bf x}} \ $ be Eulerian
coordinates of a particle and $\ {\cal D}(t)\ $ a volume of the physical
space occupied by a fluid at time $\ t$. When $\ t\ $
belongs to the finite interval $\ [t_0,t_1]\ , \ {\cal D}(t)\ $ generates a
four-dimensional domain $\ \Omega\ $ in the time-space.
A particle is labelled by its position $ \ {\bf X}\ $ in a reference space
$\ {\cal D}_0$.
For example, if $\ {\cal D}(t)\ $ consists always of the same particles $\
{\cal D}_0={\cal D}(t_0)\ $ and we can define the motion of the continuum
as a diffeomorphism from $\ {\cal D}(t_0)\ $ into $\ {\cal D}(t)$:
$$
{\bf x} ={\varphi} _t({\bf X}).
\eqno{(2.1)}
$$
The motion (2.1) of the fluid is generalized in the following parametric
form
$$
\cases{t\ =g(\lambda,{\bf X})\cr {\bf x}=\phi(\lambda,{\bf X})} \hskip
1 cm \hbox{or} \hskip 1 cm {\bf z}={\bf M}({\bf Z}),
\eqno{(2.2)}
$$
where $\ {\bf Z}=\pmatrix{\lambda\cr {\bf X}}\ $ belongs to a reference
space denoted
$\ \Omega_0$ and ${\bf M}$ is a diffeomorphism from a
reference space $\Omega_0$
into the time-space $\Omega$ occupied by the medium.

\noindent Equations (2.2) lead to the following expressions for the
differentials $\ dt \ $ and $\ d{\bf x}$,
$$
\pmatrix{ dt\cr d{\bf x}}=B\pmatrix{ d\lambda\cr d{\bf x}},
\eqno{(2.3)}
$$
where
$$
B =\ \displaystyle{\partial {\bf M}\over \partial {\bf Z}}=\left[\matrix{\
\displaystyle {\partial g\over \partial \lambda}\ \ , \ {\partial g\over
\partial {\bf X}} \cr\cr
\displaystyle{\partial
\phi \over \partial \lambda} \ ,\ {\partial \phi\over \partial {\bf X}}}
\right].
\eqno{(2.4)}
$$
In an explicit form, we obtain from (2.3)-(2.4)
$$
\cases{dt=\displaystyle{\partial g\over \partial \lambda}\ d\lambda +
{\partial g\over
\partial {\bf X}}\ d{\bf X}, \cr \cr d{\bf x}=\displaystyle{\partial
\phi \over \partial \lambda} \ d\lambda + {\partial \phi\over \partial {\bf
X}}\ d{\bf X}.}
\eqno{(2.5)}
$$
Eliminating $\ d\lambda\ $ from the first equation of (2.5) and
substituting into the second, we obtain
$$
d{\bf x}={\bf u}\ dt + F d{\bf X},
$$
where the velocity $\ {\bf u}\ $ and the deformation gradient $\ F\ $ are
defined by
$$
{\bf u}= {\partial
\phi \over \partial \lambda}\ \ \big ({\partial g\over \partial \lambda
}\big)^{-1}\ \ , \
F \ =\ {\partial \phi\over \partial {\bf X}} \ -\ \ {\partial \phi\over
\partial
\lambda}\ {\partial g\over \partial {\bf X}}\big ({\partial g\over
\partial \lambda }\big)^{-1}.
\eqno{(2.6)}
$$
Let
$$
\cases{t=G(\lambda,{\bf X},\varepsilon)\cr\cr {\bf x}=\Phi (\lambda,{\bf
X},\varepsilon)}\hskip 1 cm \hbox{ or }\hskip 1 cm {\bf z}={\bf
M}_\varepsilon({\bf Z}),
\eqno{(2.7)}
$$
where $\ \varepsilon\ $ is a scalar defined in the vicinity of zero,
be a one-parameter family of virtual motions of the medium such that
$$
{\bf M}_0({\bf Z})={\bf M}({\bf Z} ).
$$

We define the virtual displacement
$\ {\bfmat{\zeta}}=( \tau,{\bfmat{\xi}})\ $
associated with the virtual motion (2.7):
$$
\matrix{ \tau =\ \displaystyle{\partial G\over \partial
\varepsilon}(\lambda,{\bf X},0) \ , \
{\bfmat{\xi}}=\ \displaystyle{\partial \Phi\over \partial \varepsilon}
(\lambda,{\bf X},0) \cr\cr
\hbox{or},\ \ \ \ \ {\bfmat{\zeta}} =\ \displaystyle{\partial
{\bf M}_\varepsilon\over \partial \varepsilon}\ ({\bf Z)}
\vert_{\varepsilon = 0} }.
\eqno{(2.8)}
$$
From the mathematical point of view, spaces $\Omega_0$ and $\Omega$ play a
symmetric role. From the physical
point of view they are not symmetric: the tensorial quantities
(thermodynamic or mechanical) are defined either on
$\Omega_0$ or on
$\Omega$. Their image in the dual space depends on the motion of the
medium. For example, the potential of
body forces $\Pi$ is defined on $\Omega$ and the entropy
$s$ is defined on $\Omega_0$.

\noindent Let us consider any tensorial quantity represented in the form
$$
(t,{\bf x})\ \in\ \Omega \longrightarrow f(t,{\bf x}).
$$
The tensorial quantity associated with the varied motion is
$$
\buildrel\sim\over f (\lambda,{\bf X},\varepsilon)=f \Big( G(\lambda,{\bf X},
\varepsilon)
\ , \ \Phi(\lambda,{\bf X},\varepsilon)\Big)=f\Big({\bf
M}_\varepsilon({\bf Z})\Big).
$$
We define the variation of $f$ by
$$
\delta f=\ {\partial \buildrel\sim\over f \over \partial \varepsilon}\
(\lambda,{\bf X},0).
$$
For a tensorial quantity represented in the form
$$
(\lambda, {\bf X})\ \in\ \Omega_0 \longrightarrow h(\lambda, {\bf X}),
$$
the tensorial quantity associated with the varied motion is unchanged and
$$
\delta h=\ 0
$$
Let the Lagrangian of the medium be defined in the form
$$
L=L\Big({\bf z},\ {\partial{\bf M} \over \partial {\bf Z}} \ , \ {\bf
Z}\Big)=L({\bf z}, B, {\bf Z})\ .
$$
This expression contains the gas dynamics model
where the Lagrangian is [3]
$$
\matrix{L=\ {1\over 2}\ \rho\Big(1+\vert{\bf
u}\vert^2\Big)-\varepsilon(\rho,s)-\rho\Pi({\bf z})\cr \cr
=\displaystyle{1\over
2}\rho{\bf V}^*{\bf V}-\varepsilon(\rho,s)-\rho\Pi({\bf z})}.
\eqno{(2.9)}
$$
Here $\ \displaystyle{\bf V}=\pmatrix{ 1\cr{\bf u}} \ $ is the time-space
velocity,
$\displaystyle\ \Pi({\bf z})\ $ is the external force potential,
$\rho$ is the density defined by
$$
\rho\ \det\ F=\rho_0({\bf X}),
$$
and $\ s\ $ is the entropy per unit mass defined by
$$
s=s_0({\bf Z}).
$$
It is not necessary to assume that $\ s_0\ $ is a function of $\ {\bf X}\ $
only. This property will be a consequence of the variational principle
(see formula (D.1) in Appendix D).

\noindent The Hamilton action is:
$$
a=\int_{\Omega}\ L({\bf z}, B, {\bf Z})\ d\Omega.
\eqno{(2.10)}
$$
For the gas dynamics we obtain from (2.4), (2.6):
$$
\displaystyle{\bf V} = \matrix{\pmatrix{ 1\cr {\bf u}} =
\displaystyle {B {\bfmat{\ell}} \over \mu}, \hskip 0,5 cm }
\eqno{(2.11)}
$$
$$
\hbox{where}\hskip 0,25 cm
{\bfmat{\ell}}^*=(1,0,0,0) \hskip 0,25 cm \hbox {and}\ \ \
 \mu={\bfmat{\ell}}^* B {\bfmat{\ell}}= b_1^1=\
\displaystyle{\partial g \over \partial \lambda}.
$$
Consequently,
$$
\ {1\over 2}\ \Big(1+\vert{\bf u}\vert^2\Big)={1\over 2}\ \ {{\bfmat{\ell}}^*
B^*B \ {\bfmat{\ell}}\over \mu^2}.\
\eqno{(2.12)}
$$
Moreover,
$$
\rho\ =\ {\mu\over \det B}\ \rho_0({\bf X}).
\eqno{(2.13)}
$$
In the Lagrangian coordinates Hamilton's action (2.10) is
$$
\displaystyle\ a= \int_{\Omega_0}\ L({\bf z},
B,{\bf Z}) \det B \ d\Omega_0,
$$
and the varied action is
$$
a(\varepsilon)= \int_{\Omega_0}\ L \Big ( {\bf M}_\varepsilon({\bf Z}),
\ {\partial
{\bf M}_\varepsilon({\bf Z})\over \partial {\bf Z}}, \ {\bf Z}\Big )\ \det
\left( {\partial
{\bf M}_\varepsilon({\bf Z})\over \partial {\bf Z}}\right) \ d{\Omega_0}.
$$
Let $ \ T(\Omega)\ $ be the tangent bundle of $\Omega.$

{\bf The Hamilton principle} is:

{\it For any continuous virtual displacement} ${\bfmat{\zeta}} $ {\it
belonging to} $\
T(\Omega) \ $ {\it such that} $\ {\bfmat{\zeta}}=0\ $ {\it on} $\ T(\partial
\Omega),$
$$
\displaystyle\ \delta a=\ {da\over d\varepsilon}\ \vert_{\varepsilon\ =\
0}=0.
$$
Consequently,
$$
\delta a=\int_{\Omega_0}\Big\{\det B\ {\partial L\over \partial {\bf
z}}{\bfmat{\zeta}}
+\det B\ tr\Big(\ {\partial L\over \partial B}\delta B\Big) + L\
\delta(\det B)\Big\}
d\Omega_0.
$$
The Euler-Jacobi identity
$$
\matrix{ \delta\det B = tr \Big(\ \displaystyle{\partial \det B
\over \partial B}\ \ \delta
B\Big) =tr\ (B^{-1}\det B \delta B)}
=\det B\ tr (B^{-1}\ \ \delta B)
$$
and the relation issued from definitions (2.4), (2.8)
$$
\displaystyle\ \delta B=\ {\partial {\bfmat{\zeta}}\over \partial
{\bf Z}}\ =\ {\partial {\bfmat{\zeta}}\over \partial {\bf z}}\ B
$$
yield
$$
\delta a = \int_{\Omega}\left\{\matrix{{\bf S}^*{\bfmat{\zeta}}
\ +\ tr\ \pmatrix{T\ \displaystyle{\partial {\bfmat{\zeta}}\over \partial
{\bf z}}\ }} \right\} \ d\Omega,
$$
where
$$
{\bf S}^*\ =\ \ {\partial L\over \partial {\bf z}} \hskip 0,5 cm \hbox{and}
\hskip 0,5 cm \ T\ =\ L \ I \ +\ B\ {\partial L
\over
\partial B}.
$$
The Gauss-Ostrogradskii formula involves
$$
\delta a\ =\ \int_\Omega\pmatrix{{\bf S}^* \ -\
\hbox{Div }T}{\bfmat{\zeta}}\ d\ \Omega \ +\
\int_{\partial \Omega}{\bf N}^*T{\bfmat{\zeta}}\ d\omega,
$$
where $\ {\bf N}^*\ $ is the external normal to $\ \partial \Omega\ $ and
$d\omega$ is the local measure of ${\partial \Omega}$.

\noindent If the motion is continuous on $\ \Omega \ {\rm and} \
{\bfmat{\zeta}}\ =\ 0\ $ on $\
\partial \Omega\ $, we get
$$
\delta a\ =\ \int_\Omega\pmatrix{{\bf S}^* \ -\ \hbox{Div }T}
{\bfmat{\zeta}}\ d\ \Omega.
$$
Consequently, the governing equations are
$$
{\bf S}^* \ -\ \hbox{div } T\ =\ 0.
\eqno{(2.14)}
$$
In Appendix A we verify that (2.14) corresponds to classical momentum and
energy equations. If there exists a surface $\Sigma$ of discontinuity
of \ $ B \ $ separating $\ \Omega\ $ into two parts $\ \Omega_1\ $ and
$\ \Omega_2$ we
get
$$
\delta a\ =\ \int_{\Omega_1}\pmatrix{{\bf S}^* \ -\ \hbox{Div }T} \
{\bfmat{\zeta}}\ d\ \Omega_1 \
+\ \int_{\Omega_2}\pmatrix{{\bf S}^*
\ -\ \hbox{Div }T} \ {\bfmat{\zeta}}\ d\ \Omega_1 \ +\ \int_{\Sigma}{\bf
N}^*[T]{\bfmat{\zeta}}\ d\ \omega,
$$
where $\ [T]\ =\ T_1 - T_2\ $ denotes the jump of $\ T\ $ $ \Sigma$.

\noindent Consequently the fundamental lemma of calculus of variations
involves the Rankine-Hugoniot conditions
$$
{\bf N}^*[T]\ =\ 0.
\eqno{(2.15)}
$$
Because $\ {\bf N}^*\ $ is collinear to
$\ [- {D}_n \ , \ {\bf n}^*],\ $ where $\ {D}_n\ $
is the normal velocity of $\Sigma $
and $ \ {\bf n} \ $ is the normal unit space vector, relations (2.15) are
the classical Rankine-Hugoniot conditions representing the
conservation of momentum and energy through the shock (see Appendix A and
[7]).

\bigskip

\section{Two-fluid models: General calculations}

We shall study now two-fluid motions, the method being extended to any
number of components. We generalize the representation of the motion (2.2)
considering the motion of a two-fluid mixture as two
diffeomorphisms [14]
$$
{\bf z}\ =\ {\bf M}_\alpha\ ({\bf Z}_\alpha),
$$
where $\ {\bf Z}_\alpha\ =\ \left[\matrix{\lambda_\alpha\cr {\bf
X}_\alpha}\right]\ $ belongs to a reference space
$ \ \Omega_\alpha\ $ associated with the $\ \alpha-$th component.
The Jacobian matrix is defined by the formula
$$
B_\alpha\ =\ \ {\partial {\bf M}_\alpha\over \partial
{\bf Z}_\alpha}\ ({\bf Z}_\alpha).
$$
The velocity $\ {\bf u}_\alpha\ $ and the deformation gradient $\
F_\alpha\ $ are defined similarly to (2.6).

\noindent Two one-parameter families of virtual motions are associated
with the two diffeomorphisms
$$
\cases{{\bf z}\ =\ {\bf M}_{1, \varepsilon_1}\ ({\bf Z}_1)\cr
{\bf z}\ =\ {\bf M}_2\ (\bf Z_2)}
\eqno{(3.1)}
$$
such that $\displaystyle\ {\bf M}_{1,0}\ ({\bf Z}_1)\ =\ {\bf M}_{1 }\
({\bf Z}_1)$, and
$$
\cases{{\bf z}\ =\ {\bf M}_1\ ({\bf Z}_1) \cr {\bf z}\ =\ {\bf M}_{2,
\varepsilon_2}\ ({\bf Z}_2)}
\eqno{(3.2)}
$$
such that $\ \displaystyle\ {\bf M}_{2,0}\ ({\bf Z}_2)\ =\ {\bf M}_{2 }\
({\bf Z}_2)$.

\noindent The two families extend the concept of virtual motion defined in
section 2.
Consider the family (3.1), but all the consequences are
the same for the family (3.2). We define the virtual displacement of the
first component by the relation
$$
{\bfmat{\zeta}}_1\ =\ \ {\partial {\bf M}_{1,\varepsilon_1}\over \partial
\varepsilon_1} ({\bf Z}_1)
{\ \displaystyle\left\vert_{\varepsilon_1=0}\right. }.
$$
Virtual motion (3.1) generates a displacement of component "2"
$$
{\bf M}_{1, \varepsilon_1}\ ({\bf Z}_1)\ =\ {\bf M}_2\ ({\bf Z}_2),
$$
which defines $\ {\bf Z}_2\ $ as a function of $\ {\bf Z}_1\ $ and $\
\varepsilon_1$.
Taking the derivative with respect to $\ \varepsilon_1\ $ and denoting
$$
\delta_1{\bf Z}_2\ =\ \ {\partial{\bf Z}_2 \over \partial \varepsilon_1}
{\ \displaystyle\left\vert_{\varepsilon_1=0} \right. },
$$
we obtain
$$
\delta_1{\bf Z}_2\ =\ B^{-1}_{2}{\bfmat{\zeta}}_1.
\eqno{(3.3)}
$$
Let us consider any tensorial quantity $\ f\ $ in Eulerian coordinates:
$$
{\bf z}\ \in\ \Omega\ \longrightarrow\ f ({\bf z}).
$$
The tensorial quantity associated with the varied motion is then
$$
\buildrel\sim\over f ({\bf Z}_1, \varepsilon_1)\ =\ f\Big({\bf M}_{1,
\varepsilon_1} ({\bf
Z}_1)\Big)
$$
and consequently $\displaystyle\ \delta_1f\ =\ \
{\partial{\buildrel\sim\over f} \over
\partial \varepsilon_1} ( {\bf Z}_1, 0)\ $ is the variation of $\ f$.

\noindent Let us consider the Lagrangian of the medium in the representation
$$
L\ =\ L({\bf z},B_1,B_2,{\bf Z}_1, {\bf Z}_2).
\eqno{(3.4)}
$$
For example, the Lagrangian of a two-fluid mixture is [1-2, 4-6, 11]:
$$
L\ =\ \ {1\over 2}\ \rho_1\ (1 + \vert{\bf u}_1\vert ^2) \
+\ {1\over 2}\ \rho_2\ (1 + \vert{\bf u}_2\vert ^2) \ -\
W(\rho_1,\rho_2,s_1,s_2, {\bf u}_2 - {\bf u}_1 ) \ -\
\rho \ \Pi({\bf z}),
\eqno{(3.5)}
$$
where $ \ \rho_\alpha\ $ is the density of the $\alpha
-{\hbox{th}}\ $ component defined by
$$
\rho_\alpha\ \det\ F_\alpha\ =\ \rho_{0,\alpha}({\bf X}_\alpha),
$$
$\ s_{\alpha} \ $ is the entropy per unit mass of
the $\ \alpha-{\hbox{th}}\ $ component defined by the relation
$$
s_\alpha\ =\ s_{0,\alpha}({\bf Z}_\alpha),
$$
and $\ \Pi({\bf z})\ $ is the potential of external forces.

\noindent
One can easily obtain the formulae analogous to (2.11)-(2.13) for $\
{\rho}_\alpha\ $ and
$\ {\bf u}_\alpha\ $ in terms of $\ B_\alpha$ and ${\bf Z}_\alpha$.
Hence, the Lagrangian (3.5) may be rewritten in the form (3.4).
The variation associated with the application (3.1) yields
$$
\delta_1 a\ =\ \int_{\Omega_1}\delta_1\ (L\det B_1)\ d\Omega_1.
$$
Calculations presented in Appendix B give the following result:
$$
\delta_1a\ =\ \int_{\Omega}\left\{\matrix{tr\left(T\ \displaystyle{\partial
{\bfmat{\zeta}}_1\over \partial {\bf z}}-T_1\ {\partial
\delta_1{\bf Z}_2\over \partial {\bf z}}\right) \ +\ {\bf
S}_1^*{\bfmat{\zeta}}_1 }\right\}\ d\Omega,
\eqno{(3.6)}
$$
where
$$
\cases{T\ =\ L\ I +B_1\ \displaystyle{\partial L\over \partial B_1}
+ B_2\ {\partial L\over \partial B_2}, \cr\cr
T_1\ =\ B_2\
\displaystyle {\partial L\over \partial B_2} B_2, \cr \cr
{\bf S}^*_1\ =\
\displaystyle\ {\partial L\over \partial {\bf
Z}_2}B_2^{-1}+{\partial L\over \partial{\bf z}}. \ }
\eqno{(3.7)}
$$

The Gauss-Ostrogradskii formula and relation (3.3) involve
$$
\delta_1a\ =\ \int_{\Omega}\Bigg\{ {\bf S}_1^*- \hbox{Div }
T+(\hbox{Div } T_1)B_2^{-1} \Bigg\}\
{\bfmat{\zeta}}_1\ d\Omega
+ \int_{\partial \Omega}{\bf N}^*(T-T_1B_2^{-1})\ {\bfmat{\zeta}}_1\
d\Sigma.
\eqno{(3.8)}
$$
Using the arguments described in section 2, we obtain from (3.8) both
governing equations and Rankine-Hugoniot conditions for component ''1''
$$
{\bf S}_1^*- \hbox{Div } T+(\hbox{Div } T_1)B_2^{-1}=0.
\eqno{(3.9)}
$$
$$
{\bf N}^* [T-T_1B_2^{-1}]\ =\ 0.
\eqno{(3.10)}
$$
Since $\displaystyle\ T-T_1B_2^{-1}\ =\ L \ I +B_1\ {\partial L\over
\partial B_1},\ $
equation (3.10) is equivalent to
$$
{\bf N}^*\left[\matrix{L\ I +B_1\ \displaystyle {\partial L\over
\partial B_1}\ }\right]\ =\ 0.
\eqno{(3.11)}
$$
Equations for component ''2'' are obtained by permutation indexes ''1'' and
''2'':
$$
{\bf S}^*_2-\hbox{Div }T+(\hbox{Div }T_2)B^{-1}_1\ =\ 0
\eqno{(3.9')}
$$
and
$$
{\bf N}^* [T-T_2B_1^{-1}]\ =\ 0
\eqno{(3.10')}
$$
Formula (3.10') can be rewritten in an equivalent form
$$
{\bf N}^*\left[\matrix{L\ I +B_2\ \displaystyle {\partial L\over
\partial B_2}\ } \right]\ =\ 0.
\eqno{(3.11')}
$$
Let us remark that the equations (3.9) and (3.9') are not in a divergence
form. Let us denote
$$
{\bf S}^*\ =\ \ {\partial L\over \partial {\bf z}}\
\eqno{(3.12)}
$$
and
$$
{\bf S}^*_{0\alpha}\ =\ \det \ B_\alpha\ {\partial L\over \partial {\bf
Z}_\alpha} \ , \ T_{0\alpha}\ =\ -\det B_\alpha {\partial L\over
\partial B_\alpha} B_\alpha.
\eqno{(3.13)}
$$
Identities
$$
T_{02}\equiv -(\det B_2)B^{-1}_2T_1 \ , \ T_{01}\equiv -(\det
B_1)B^{-1}_1T_2,
$$
$$
{\bf S}^*_1 \equiv {\bf S}^*+\ {1\over \det B_2}\ {\bf S}^*_{0
2}B^{-1}_2 \ , \ {\bf S}^*_2 \equiv{\bf S}^*+\ {1\over \det B_1}\ {\bf
S}^*_{0 1}B^{-1}_1,
$$
and (1.1) involve that (3.9) and (3.9') are equivalent to
$$
{\bf S}^*- \hbox{Div } T+\displaystyle\ {1\over \det B_1}\ ({\bf
S}^*_{01}-\hbox{Div}_1\ T_{01})B^{-1}_1\ =\ 0,
\eqno{(3.14)}
$$
and
$$
{\bf S}^*- \hbox{Div } T+\displaystyle\ {1\over \det B_2}\
({\bf S}^*_{02}-\hbox{Div}_2\ T_{02})B^{-1}_2\ =\ 0,
\eqno{(3.14')}
$$
where $\ \hbox{Div}_{\alpha}\ T_{0\alpha}\ $ means the
divergence of $\ T_{0\alpha}\ $ with respect to the $\ \alpha-$th
Lagrangian coordinates.
The following identity $\ (C.1)\ $ proved in Appendix C,
$$
{\bf S}^*- \hbox{Div }
T+\displaystyle\ {1\over \det B_1}\ ({\bf S}^*_{01}-
\hbox{Div}_1\ T_{01})B^{-1}_1 \ +\displaystyle\ {1\over \det B_2}\
({\bf S}^*_{02}-\hbox{Div}_2\ T_{02})B^{-1}_2\equiv 0
$$
and (3.14), (3.14') yield
$$
{\bf S}^* - \hbox{Div } T\ =\ 0.
\eqno{(3.15)}
$$
We see in the next section that for fluid mixtures, equation (3.15)
represents the conservation laws of total momentum and total energy.
Hence, (3.14) and (3.14') are equivalent to:
$$
{\bf S}^*_{01} - \hbox{Div}_1 T_{01}\ =\ 0,
\eqno{(3.16)}
$$
$$
{\bf S}^*_{02} - \hbox{Div}_2 T_{02}\ =\ 0.
\eqno{(3.16')}
$$
Equations (3.16) and (3.16') represent equations of motion for each
component of a two-fluid medium in Lagrangian coordinates. They are less
useful than
equations in the Eulerian coordinates. However, they are in
a divergence form and involve the conservation of the total
momentum and the total energy.

\bigskip

\section{Application to two-fluid mixtures}

For a two-fluid mixture, the Lagrangian is (see (3.5) ) :
$$
L\ =\ {1\over 2}\rho_1\vert {\bf V}_1\vert ^2+\
{1\over 2}\rho_2\vert {\bf V}_2\vert
^2- W (\rho_1,\rho_2,s_1,s_2,{\bf w})-\rho\Pi({\bf z}),
\eqno{(4.1)}
$$
where
$$
{\bf V}_\alpha\ =\ \pmatrix{1\cr {\bf u}_\alpha}\equiv\ {B_\alpha\over
\mu_\alpha}\
{\bfmat{\ell}} \ , \ \pmatrix{0\cr {\bf w}}\ =\ {\bf V}_2-{\bf V}_1.
\eqno{(4.2)}
$$
$$
\rho_\alpha\ =\ \displaystyle\ {\mu_\alpha\
\rho_{0\alpha}\ ({\bf X}_\alpha)\over \det B_\alpha},\
\eqno{(4.3)}
$$
$$
s_\alpha\ =\ s_{0\alpha}({\bf Z}_\alpha).
\eqno{(4.4)}
$$
Definitions (4.1)-(4.4) involve the governing equations (3.9),
(3.9') in the following form (see Appendix D):
$$
\cases{\ \displaystyle{\partial {\bf K}_\alpha\over \partial t}+
\hbox{rot }
{\bf K}_\alpha \times \ {\bf u}_\alpha=\nabla^*( {\bf R}_\alpha- {\bf
K}_\alpha^* {\bf u}_\alpha)+\theta_\alpha\nabla^*s_\alpha,\cr\cr
\displaystyle {\partial \rho_{\alpha}\over \partial t} +\hbox{div
}(\rho_\alpha{\bf u}_\alpha)\
=\ 0,\cr\cr \displaystyle {\partial (\rho_{\alpha} s_{\alpha})\over
\partial t} +\hbox{div }\displaystyle(
\rho_\alpha s_\alpha{\bf u}_\alpha)\ =\ 0,}
\eqno{(4.5)}
$$
where
$$
\cases{\rho_\alpha{\bf K}^*_\alpha\equiv\
\displaystyle{\partial L\over \partial {\bf u}_\alpha}= \rho_\alpha {\bf
u}_\alpha^*-(-1)^\alpha\ {\partial W\over
\partial {\bf w}} \ , \ \ \ \hbox{with} \ \ {\bf w}={\bf u}_2-{\bf
u}_1, \cr\cr R_\alpha
\equiv\ \displaystyle{\partial L\over \partial \rho_\alpha}={1\over
2}\vert{\bf u}_\alpha\vert ^2-
{\partial W\over \partial \rho_\alpha}-\Pi ,\cr\cr
\rho_\alpha\theta_\alpha\equiv
-\displaystyle{\partial L\over \partial {s}_\alpha}={\partial W\over \partial
{s}_\alpha}.}
$$
We notice here that the governing equations were obtained earlier by a
different method in [11].
Equations (4.5) yield the conservation laws for the total momentum and the
total energy associated with (3.15)
$$
\displaystyle\ {\partial \over \partial t}
(\rho_1{\bf K}_1^*+\rho_2 {\bf K}_2^*) +
$$
$$
+ \hbox{div }\Big(\rho_1{\bf u}_1{\bf K}_1^*+\rho_2{\bf u}_2{\bf
K}_2^*+(\rho_1\ {\partial W\over \partial \rho_2}+\rho_2
{\partial W\over \partial \rho_2}-W) \ I \Big) +\rho\nabla \Pi=0,
$$
$$
\displaystyle
{\partial \over \partial t}\Big ( {\displaystyle\sum
^{2}_{\alpha=1}\rho_\alpha\ {1\over
2}\vert {\bf u}_\alpha\vert ^2+\rho\Pi+ W - {\partial W\over\partial {\bf
w}}{\bf w} }\Big ) +
$$
$$
+\hbox{div }
\Big({\displaystyle \sum ^{2}_{\alpha=1}
\rho_\alpha{\bf u}_\alpha({\bf K}_\alpha^*{\bf u}_\alpha-R_\alpha)\Big)
- \rho\ {\partial \Pi\over \partial t}\ } = 0.
\eqno{(4.6)}
$$
In the general case, they are the only conservation laws admitted by the
system. Hence, the system (4.5) is not in a divergence form.
The Rankine -Hugoniot conditions (3.10)-(3.10') (or (3.11)-(3.11')) for
this system are obtained in
Appendix A(b) (see formulae (A.9)); they are
$$
\bigg[ - D_n(L-\rho_\alpha{\bf K}^*_\alpha{\bf u}_\alpha) \ +\
 \rho_\alpha {\bf n}^* \ {\bf u}_\alpha
(R_\alpha - {\bf K}^*_\alpha{\bf u}_\alpha) \bigg] \ =\ 0,
\eqno{(4.7^1)}
$$
$$
\bigg[ - D_n \rho_\alpha{\bf K}^*_\alpha+ {\bf n}^* \Big (
(L -\rho_\alpha {R_\alpha}) \ I +
\rho_\alpha {\bf u}_\alpha {\bf K}^*_\alpha
\Big )\bigg]
\ =\ 0.
\eqno{(4.7^2)}
$$
In addition, the mass conservations laws are expressed in the form
$$
\bigg[ \rho_\alpha ( {\bf n}^* {\bf u}_\alpha-D_n)\bigg] \ =\ 0.
\eqno{(4.7^3)}
$$
Let us consider {\bf shock waves} when
$\ {\bf n}^* {\bf u}_\alpha-D_n\ \ne\ 0 $.
Taking into account $(4.7^2)$ and $(4.7^3)$ we obtain
$$
\bigg[ {\bf K}_\alpha-
({\bf K}^*_\alpha {\bf n} ){\bf n} \bigg] \ =\ 0,
\eqno{(4.8^1)}
$$
which means that $\ [{\bf K}_\alpha]\ $ is normal to the shock.
By using $(4.7^1)$ and the identity
$$
\bigg[ L-\rho_\alpha {\bf K}^*_\alpha {\bf u}_\alpha
\bigg]\ =\ \ {1\over D_n}\ {\bf n}^* \bigg[ \rho_\alpha
{\bf u}_\alpha (R_\alpha - {\bf K}^*_\alpha {\bf u}_\alpha ) \bigg],
$$
we obtain from relation $(4.7^2)$
$$
\bigg[ {\bf K}^*_\alpha {\bf u}_\alpha -R_\alpha - D_n( {\bf
K}^*_\alpha {\bf n} ) \bigg] \ =\ 0.
\eqno{(4.8^2)}
$$
Consequently, $(4.7^1)$ yields to
$$
\bigg[ L-\rho_\alpha R_\alpha+\rho_\alpha ( {\bf n}^* {\bf u}_\alpha -D_n )
{\bf K}^*_\alpha {\bf n} \bigg] \ =\ 0.
\eqno{(4.8^3)}
$$
Equations $(4.7^3)$, $(4.8^1)$ - $(4.8^3)$ form a complete set
of eight scalar
Rankine-Hugoniot conditions, representing the conservation of mass,
momentum and energy of $\ \alpha $-th component.
For the gas dynamics, these conditions correspond to the classical shock
conditions. We have to emphasize here that
for the two-fluid model equations $(4.8^{1}) - (4.8^{3})$
do not involve the conservation of the total momentum
and the total energy through the shock.
We notice that the jump conditions $(4.8^{1}) - (4.8^{2})$ were obtained
earlier for
barotropic fluids using the variations in the Eulerian coordinates [1-2].

\noindent For {\bf contact discontinuities}, when $\ {\bf n}^* {\bf
u}_\alpha-D_n\ = \ 0 $, we get from $(4.7^1)-(4.7^3)$
$$
\bigg[ L-\rho_\alpha R_\alpha \bigg] \ =\ 0.
\eqno{(4.8^4)}
$$
For the gas dynamics equation $(4.8^4)$ corresponds to the continuity of
the pressure. All the jump
conditions are obtained from the Hamilton principle without any ambiguity.

\bigskip

\section{Conclusion}

Using the Hamilton principle we have obtained in the general conservative
case
the governing equations for two-fluid mixtures (4.5). The equations for
the total
quantities ( total momentum and total energy) are in a divergence form
(see (4.6)). The equations for the
components are in divergence form only in the Lagrangian coordinates (see
(3.16)-(3.16')).
The Hamilton principle gives also a set of Rankine-Hugoniot conditions
$(4.7^1) -(4.7^2)$.
Together with the equations of mass $(4.7^3)$ they form a complete set of
the jump relations. For the gas dynamics model
they coincide with the classical Rankine-Hugoniot conditions of
conservation of mass, momentum and energy.

\bigskip

\section{Appendix A}

In gas dynamics, the Lagrangian (2.9) is a function of $\ {\bf V}\ =
\ \pmatrix{1\cr {\bf u}},\ \rho, \ s\ $ and $\ {\bf z}$.
Formulae (2.11)-(2.13) allow us to consider the Lagrangian
as a function of $\ {\bf z}, B, {\bf Z}$:
$$
\matrix{{\bf V}\ =\ \ \displaystyle {B\over \mu}\ {\bfmat{\ell}} \ , \
\mu \ =\
{\bfmat{\ell}}^* B {\bfmat{\ell}}, \cr\cr \rho\ \det\ B\ =\
\mu \ \rho_0({\bf X}), \ s\ =\ s_0({\bf Z})}.
\eqno{(A.1)}
$$
We need to calculate $\displaystyle\ T\ =\ L\ I +B\ {\partial
L\over \partial B}\ $. From (A.1) we have:
$$
d\mu\ =\ {\bfmat{\ell}}^*\ dB\ {\bfmat{\ell}}\ =\ tr
({\bfmat{\ell}}{\bfmat{\ell}}^*\ dB),
\eqno{(A.2)}
$$
$$
d{\bf V}\ =\ \ {dB\over \mu}\ {\bfmat{\ell}}-\ {B{\bfmat{\ell}}\over \mu^2}\
d\mu\ =\
{dB\over \mu}\ {\bfmat{\ell}}-\ {B{\bfmat{\ell}}\over \mu^2}\ tr\
({\bfmat{\ell}}{\bfmat{\ell}}^*dB),
\eqno{(A.3)}
$$
$$
{\partial \rho\over \partial B}\ =\ \rho_0({\bf X}) \mu
{\partial \over \partial B}
\pmatrix{\displaystyle{1\over \det B}} + \rho_0({\bf X}) {1\over \det B}
{\partial \mu\over \partial B}\ =\
\rho\pmatrix{\displaystyle\ {{\bfmat{\ell}}{\bfmat{\ell}}^*\over \mu}-B^{-1}}.
\eqno{(A.4)}
$$
Moreover,
$$
{\partial s\over \partial B}\ =\ 0.
\eqno{(A.5)}
$$
The differential of $\displaystyle\ L\ =\ L({\bf V}, \rho, s,{\bf z})\ $ is:
$$
dL\ =\ \ {\partial L\over \partial {\bf V}}\ d{\bf V} + {\partial
L\over \partial\rho}\
d\rho + {\partial L\over \partial s}\ ds+ {\partial L\over \partial {\bf
z}}\ d{\bf z}.
$$
Taking into account relations(A.2)-(A.5), we obtain
$$
dL\ =\ \ {\partial L\over \partial {\bf V}}\pmatrix{\ \displaystyle{dB
\ {\bfmat{\ell}}\over \mu} - {B {\bfmat{\ell}}\over \mu^2}\
tr({\bfmat{\ell}}{\bfmat{\ell}}^* dB) } \ +
$$
$$
+\ \ {\partial L\over \partial \rho}\ d\rho+ {\partial L\over \partial s}\
ds+ {\partial L\over \partial {\bf z}}\ d{\bf z} \ \ =
$$
$$
tr\ \pmatrix{\displaystyle{{\bfmat{\ell}}\over \mu} {\partial L\over
\partial {\bf V}}
dB- {\displaystyle{\partial L\over \partial {\bf V}}B{\bfmat{\ell}}\over
\mu^2} {\bfmat{\ell}}{\bfmat{\ell}}^* dB} +
$$
$$
+ {\partial L\over \partial \rho}tr \pmatrix{\displaystyle{\partial
\rho\over \partial
B} \ dB } + \pmatrix{\displaystyle{\partial L\over
\partial \rho}{\partial \rho\over \partial {\bf Z}}+{\partial L\over
\partial s}{\partial s\over \partial {\bf Z}}}d{\bf Z}+{\partial L\over
\partial {\bf z}}d{\bf z}.
$$
Hence,
$$
{\partial L\over \partial B}\ ={{\bfmat{\ell}} \displaystyle {\partial
L\over \partial {\bf V}}\over \mu } - {\displaystyle {{\bfmat{\ell}}
{\bfmat{\ell}}^* \big ( \displaystyle { \partial L\over \partial {\bf
V}}B{\bfmat{\ell} \big )}\over
\mu^2 } }+{\partial L\over \partial \rho}{\partial \rho\over \partial B}.
\eqno{(A.6)}
$$
Then
$$
B{\partial L\over \partial B}={B{\bfmat{\ell}} \displaystyle {\partial
L\over \partial {\bf V}}\over \mu } -{\pmatrix{\big (\displaystyle{ \partial
L\over \partial {\bf V}}B {\bfmat{\ell}}\big) B
{\bfmat{\ell}}{\bfmat{\ell}}^*}\over \mu^2}+ \rho {\partial L\over
\partial \rho}\pmatrix{\displaystyle
B{\displaystyle{{\bfmat{\ell}}{\bfmat{\ell}}^*}\over \mu}- \ I}.
$$
Taking into account (A.1), we obtain
$$
B{\partial L \over \partial B}={\bf V}{\partial
L\over \partial {\bf V}}-\pmatrix{\displaystyle{\partial
L\over \partial {\bf V}}{\bf V}}{\bf V}{\bfmat{\ell}}^*
+\rho{\partial L\over \partial \rho}({\bf V}{\bfmat{\ell}}^* -\ I)
\eqno{(A.7)}
$$
and, (A.7) yields then
$$
T=L\ I +B{\partial L\over \partial B}=\pmatrix{L-\rho
\displaystyle{\partial L\over \partial \rho}}\ I
+{\bf V}{\partial L\over \partial {\bf V}}
- \pmatrix{\displaystyle{\partial L\over \partial {\bf V}}
{\bf V}-\rho{\partial L\over \partial \rho}} {\bf V}{\bfmat{\ell}}^*.
\eqno{(A.8)}
$$
{ a) In the case} $\displaystyle L={1\over 2}\rho{\bf V}^*{\bf V}-
\varepsilon (\rho, s)-\rho \Pi$ formula (A.8) can be rewritten in the form
$$
T= p \ I +\rho{\bf V}{\bf V}^*- \pmatrix
{\displaystyle{1\over 2}{\bf V}^*{\bf V} +
\varepsilon'_\rho + \Pi } \rho{\bf V}{\bfmat{\ell}}^*,
$$
where $\ p=\rho\varepsilon'_\rho-\varepsilon$. \ \ Then,
$$
\displaystyle T=\left[\matrix{-E\ , \ \rho{\bf u}^*\cr \cr -(E+p){\bf u}
\ , \ p\ I +\rho{\bf u}{\bf u}^*}\right],
$$
where $\displaystyle E={1\over 2}\rho{\bf V}^*{\bf V}+
\varepsilon+\rho \Pi \ $ is the total volume energy.

\noindent
Moreover, since $\displaystyle \ {\bf S}^*= - \rho{\partial\Pi\over
\partial{\bf z}}$,
we get the energy and momentum equations for gas dynamics motions
$$
\cases{\displaystyle{\partial E\over \partial t}+\hbox{ div }
\Big((E+p){\bf u}\Big)-\rho\displaystyle{\partial \Pi\over
\partial t}=0, \cr \cr \displaystyle{\partial\rho{\bf u}^*\over\partial t}
+\hbox{ div } (p \ I +\rho{\bf u}{\bf u}^*)+\rho{\partial \Pi\over
\partial {\bf x}} =0.}
$$
Because ${\bf N^*}$ is collinear to $(\ -D_n,\ \ {\bf n}^*\ )$, the
classical Rankine-Hugoniot conditions (2.15)
are
$$
\Big[D_n E-(E+p){\bf n}^*{\bf u}\Big] \ = \ 0,
$$
$$
\Big[D_n \rho \ {\bf u}^*-{\bf n}^*( p\ I +\rho{\bf u}
{\bf u}^*)\Big] \ = \ 0.
$$

\noindent b) For the two-fluid model the Lagrangian is
$$
\displaystyle {L={1\over 2}\rho_1{\bf V}_1^*{\bf V}_1
+{1\over 2}\rho_2{\bf V}_2^*{\bf V}_2-W(\rho_1, \rho_2, s_1, s_2, {\bf w})}
$$
with
$$
\displaystyle {
\displaystyle\ \pmatrix{0\cr {\bf w}} = \pmatrix{0\cr
{\bf u}_2 - {\bf u}_1} ={\bf V}_2-{\bf V}_1. }
$$
We obtain similarly to (A.8)
$$
L \ I
+B_\alpha{\partial L\over \partial B_\alpha}=
\pmatrix{L-\rho_\alpha\displaystyle{\partial L\over \partial
\rho_\alpha}}\ I
+{\bf V}_\alpha {\partial L\over \partial{\bf V}_\alpha}
-\pmatrix{\displaystyle{\partial L\over
\partial{\bf V}_\alpha}{\bf V}_\alpha-\rho_\alpha{\partial L\over
\partial\rho_\alpha}} {\bf V}_\alpha{\bfmat{\ell}}^*.
$$
Here
$$
\displaystyle\ {\partial L\over \partial{\bf V}_\alpha}=\rho_\alpha{\bf
V}_\alpha^*- {\partial W\over \partial{\bf V}_{\alpha}} =
\rho_\alpha{\bf V}_\alpha^*- {\partial W\over \partial{\bf w}}
{\partial {\bf w}\over \partial{\bf V}_{\alpha}}
\equiv \rho_\alpha({ 1},{\bf K}_\alpha^*),
$$
$$
{\partial L\over \partial\rho_\alpha}= {1\over2}{\bf V}_\alpha^*{\bf
V}_\alpha-{\partial W\over \partial \rho_\alpha}-\Pi\equiv R_\alpha.
$$
In a matrix form, we have
$$
L \ I +B_\alpha{\partial L\over \partial B_\alpha} \ \ = \ \
\left[\matrix{L-\rho_\alpha{\bf K}_\alpha^*{\bf u}_\alpha\ ,
\ \rho_\alpha{\bf K}_\alpha^* \cr\cr
-\rho_\alpha{\bf u}_\alpha{\bf K}_\alpha^* {\bf u}_\alpha+\rho_\alpha
R_\alpha {\bf u}_\alpha\ , \
(L-\rho_\alpha R_\alpha)\ I + \rho_\alpha{\bf u}_\alpha{\bf
K}_\alpha^* }\right],
$$
and the Rankine-Hugoniot conditions for the two-fluid model are
$$
{\bf N}^*\left[\matrix{L-\rho_\alpha{\bf K}_\alpha^*{\bf u}_\alpha\ , \
\rho_\alpha{\bf K}_\alpha^* \cr \cr
\rho_\alpha{\bf u}_\alpha(R_\alpha-{\bf K}_\alpha^* {\bf u}_\alpha)\ , \
(L-\rho_\alpha R_\alpha)\ I + \rho_\alpha{\bf u}_\alpha{\bf
K}_\alpha^* }\right]\ = \ 0.
\eqno{(A.9)}
$$

\bigskip

\section{Appendix B}

The variation $\ \delta_1 a\ $ is:
$$
\delta_1 a=\int_\Omega\Bigg\{\delta_1 L+L \ \delta_1(\det B_1)(\det
B_1)^{-1}\Bigg\} d\Omega = \int_\Omega\Bigg\{\delta_1
L+tr\pmatrix{L\displaystyle{\partial{\bfmat{\zeta}}_1\over
\partial z}}\Bigg\}d\Omega.
$$
Formula (3.3) involves
$$
\delta_1L=tr \Bigg\{{\partial L\over \partial B_1}\delta_1B_1
+ {\partial L\over \partial B_2} \delta_1 B_2\Bigg\}
+{\partial L\over \partial{\bf Z}_2}B_2^{-1}{\bfmat{\zeta}}_1
+{\partial L\over \partial{\bf z}}{\bfmat{\zeta}}_1.
\eqno{(B.1)}
$$
By definition,
$$
\delta_1 B_1={\partial{\bfmat{\zeta}}_1\over \partial{\bf z}} B_1.
\eqno{(B.2)}
$$
Let us calculate $ \ \delta_1B_2$. We get
$$
\delta_1B_2=\delta_1\pmatrix{\displaystyle{\partial{\bf z}\over
\partial{\bf Z}_2}}\ =\
\delta_1\pmatrix{\displaystyle{\partial{\bf z}\over \partial{\bf
Z}_1}{\partial{\bf Z}_1\over \partial{\bf Z}_2}}
=\delta_1\pmatrix{\displaystyle{\partial{\bf z}\over \partial{\bf
Z}_1}}{\partial{\bf Z}_1\over \partial{\bf
Z}_2}+{\partial{\bf z}\over \partial{\bf
Z}_1}\delta_1\pmatrix{\displaystyle{\partial{\bf Z}_1\over \partial{\bf
Z}_2}}.
\eqno{(B.3)}
$$
Since for any linear mapping $\ \displaystyle A:
\delta(A^{-1})=-A^{-1}\delta A \ A^{-1},\ $ we have
$$
\delta_1\pmatrix{\displaystyle{\partial{\bf Z}_1\over
\partial{\bf Z}_2}}=-{\partial{\bf Z}_1\over \partial{\bf
Z}_2}\delta_1\pmatrix{\displaystyle{\partial{\bf Z}_2\over
\partial{\bf Z}_1}}{\partial{\bf Z}_1\over \partial{\bf
Z}_2}=-{\partial{\bf Z}_1\over \partial{\bf Z}_2}
{\partial\delta_1{\bf Z}_2\over \partial{\bf Z}_1}
{\partial{\bf
Z}_1\over \partial{\bf Z}_2}.
\eqno{(B.4)}
$$
Formulae (B.3) and (B.4) involve that
$$
\delta_1B_2= {\partial{\bfmat{\zeta}}_1\over \partial{\bf z}} B_2 -
{\partial{\bf z}\over \partial{\bf Z}_2}
{\partial\delta_1{\bf Z}_2\over \partial{\bf Z}_2} =
{\partial{\bfmat{\zeta}}_1\over \partial{\bf z}} B_2 - B_2
{\partial\delta_1{\bf Z}_2\over \partial {\bf z}} B_2.
\eqno{(B.5)}
$$
Substituting (B.2) and (B.5) into (B.1) we get
$$
\delta_1L=tr\pmatrix{\displaystyle{\partial
L\over \partial B_1}{\partial{\bfmat{\zeta}}_1
\over \partial{\bf
z}}B_1+{\partial L\over \partial B_2}\pmatrix
{\displaystyle{\partial{\bfmat{\zeta}}_1\over \partial{\bf
z}}B_2-B_2{\partial\delta_1{\bf Z}_2\over \partial{\bf z}}B_2}}+{\partial
L\over \partial{\bf Z}_2}
B_2^{-1}{\bfmat{\zeta}}_1+{\partial L \over \partial {\bf
z}}{\bfmat{\zeta}}_1=
$$
$$
= tr\Bigg\{\matrix{\pmatrix{B_1\displaystyle\ {\partial L\over \partial
B_1}+B_2\displaystyle\ {\partial L\over \partial B_2}}\
\displaystyle{\partial
{\bfmat{\zeta}}_1\over \partial {\bf z}}- B_2{\partial L\over \partial
B_2}B_2\ {\partial \delta_1{\bf Z}_2\over \partial {\bf z}}}\Bigg\}
+\displaystyle\
{\partial L\over \partial {\bf Z}_2}B_2^{-1}{\bfmat{\zeta}}_1+\ {\partial
L\over \partial {\bf z}}{\bfmat{\zeta}}_1.
$$
Hence,
$$
\delta_1 L+tr\pmatrix{\displaystyle L{\partial {\bfmat{\zeta}}_1\over
\partial {\bf z}}} \ =\ tr\pmatrix{T\displaystyle\ {\partial
{\bfmat{\zeta}}_1\over
\partial {\bf z}}-T_1\ {\partial \delta_1{\bf Z}_2\over \partial {\bf
z}}}+{\bf S}^*_1{\bfmat{\zeta}}_1,
$$
where $\ T \ , \ T_1\ $ and $\ {\bf S}_1^*\ $ are defined in (3.7).
The relation (3.6) is proved.

\bigskip

\section{Appendix C}

{\bf Theorem}: {\it The following expression is an identity}
$$
{\bf S}^*-\hbox{Div } T+\ {1\over \det B_1}\ ({\bf S}^*_{01}-\hbox{Div}
T_{01})\ B^{-1}_1+
{1\over \det B_2}\ ({\bf S}^*_{02}-\hbox{Div } T_{02})\ B^{-1}_2 \equiv 0,
\eqno{(C.1)}
$$
{\it where (see formulae (3.6), (3.12) -(3.13))}
$$
\cases{{\bf S}^*\ =\ \displaystyle\ {\partial L\over \partial
{\bf z}}, \cr\cr T =\ L\ I
+B_1\displaystyle\ {\partial L\over \partial B_1}+
B_2 {\partial L\over \partial B_2}, \cr \cr
{\bf S}^*_{0\alpha}\ = \det B_\alpha\displaystyle\
{\partial L\over \partial {\bf Z}_\alpha}, \cr\cr
T_{0\alpha}\ =\ -\det B_\alpha\displaystyle\
{\partial L\over \partial B_\alpha}B_\alpha. \cr\cr }
$$

{\bf Proof:}
Using (1.1), we get
$$
\sum^{2}_{\alpha = 1}\ {1\over \det B_\alpha}({\bf S}^*_{0\alpha}-
\hbox{Div}_\alpha T_{0\alpha})B^{-1}_\alpha\equiv \sum^{2}_{\alpha =
1}\pmatrix{\displaystyle\ {\partial L\over \partial {\bf Z}_\alpha}
+ \hbox{Div} ( B_\alpha\displaystyle\
{\partial L\over \partial B_\alpha}B_\alpha)} B_\alpha^{-1}.
$$
We have to prove that this expression is identical to
$$
\hbox{Div} \ T {\bf }-{\bf S}^*\equiv \hbox{Div}\pmatrix{L \
I +B_1\displaystyle\
{\partial L\over \partial B_1} + B_2\ {\partial L\over \partial B_2}}-\
{\partial L\over \partial {\bf z}}.\
$$
Indeed, by definition
$$
\hbox{Grad} \ L \equiv{\partial L\over \partial {\bf z}}+ {\partial
L\over\partial {\bf
Z}_1}{\partial {\bf Z}_1\over \partial {\bf z}}+{\partial L\over\partial
{\bf Z}_2}{\partial
{\bf Z}_2\over \partial {\bf z}}+tr\pmatrix{\displaystyle{\partial
L\over\partial B_1}{\partial
B_1\over \partial {\bf z}}}+tr\pmatrix{\displaystyle{\partial
L\over\partial B_2}{\partial B_2\over \partial {\bf z}}},
$$
where
$$
\pmatrix{tr\pmatrix{\displaystyle\ {\partial L\over \partial B_\alpha}\
{\partial B_\alpha\over \partial {\bf z}}}}_p\equiv\pmatrix{\displaystyle\
{\partial
L\over \partial B_\alpha}}^j_i\ {\partial B_j^i\over \partial z^p}.\
$$
Hence, we have to verify only the formula
$$
\hbox{Div }\pmatrix{B_\alpha\displaystyle\ {\partial L\over \partial
B_\alpha}B_\alpha } B^{-1}_\alpha\equiv \hbox{Div
}\pmatrix{B_\alpha\displaystyle\ {\partial L\over \partial B_\alpha}}+
tr \pmatrix{\displaystyle\ {\partial L\over \partial B_\alpha}\ {\partial
B_\alpha\over \partial {\bf z}}}.
\eqno{(C.2)}
$$
To prove the identity (C.2) we may consider only $\ L\ =\ L(B),\ $ with
$\displaystyle\ B\ =\ {\partial {\bf z}\over \partial {\bf Z}}$.
Let $\displaystyle\ A\ =\ {\partial L\over \partial B},\ $
then
$$
\ {\partial L\over\partial {z}^m}=tr\pmatrix{\displaystyle {\partial
L\over \partial B}\
{\partial B\over \partial {z}^m}}=a^i_j\ {\partial b^j_i\over
\partial {z}^m}.\
$$
Hence,

$\displaystyle\ {\partial ( b^i_j \ a^j_k \ b^k_s) \over \partial {z}^i} \
\bar {b}^s_p=
\displaystyle\ {\partial ( b^i_j \ \ \ a^j_k)\over \partial {z}^i} \
b^k_s\ \bar {b}^s_p+
b^i_j \ a^j_k\ {\partial b^k_s \over \partial {z}^i} \ \bar {b}^s_p = $

$\displaystyle\ {\partial( b^i_j \ a^j_p) \over \partial {z}^i} + b^i_j
\ a^j_k\ {\partial
b^k_s \over \partial {Z}^m}\ {\partial {Z}^m \over \partial {z}^i} \ \bar
{b}^s_p =
\displaystyle\ {\partial ( b^i_j \ a^j_p)\over \partial {z}^i} \ + b^i_j \
\bar {b}^m_i \ a^j_k\ {\partial b^k_s\over \partial {Z}^m} \bar {b}^s_p = $

$\displaystyle\ {\partial ( b^i_j \ a^j_p)\over \partial {z}^i} \ +
a^m_k\ {\partial b^k_m\over \partial {Z}^s} \bar
{b}^s_p = \displaystyle\ {\partial ( b^i_j \ a^j_p)\over \partial
{z}^i} \ + a^m_k\ {\partial b^k_m\over \partial
{z}^l} \ b^l_s \bar {b}^s_p = \displaystyle\
{\partial( b^i_j \ a^j_p) \over \partial {z}^i} \ + a^m_k\
{\partial b^k_m\over \partial {z}^p}. $

This is a proof of (C.2) and consequently (C.1).

\bigskip

\section{Appendix D}

First of all, we obtain the governing equations for each
component in the Lagrangian coordinates. These equations yield easily the
governing equation in the Eulerian
coordinates. In the Lagrangian coordinates equations $(3.16) - (3.16^{' })$
are
$$
{\bf S}_{0\alpha}^* -\hbox{Div}_\alpha T_{0\alpha}^*=0,
$$
where
$$
{\bf S}^*_{0\alpha}=\det B_\alpha\ {\partial L\over \partial {\bf
Z}_\alpha}\ = R_\alpha\mu_\alpha
\ {\partial \rho_{0\alpha}\over \partial {\bf
Z}_\alpha} - det B_\alpha \ \rho_\alpha\theta_\alpha\\ {\partial
s_{0\alpha}\over \partial {\bf Z}_\alpha},
$$
and
$$
T_{0\alpha}=-\det B_\alpha\ {\partial L\over \partial B_\alpha}B_\alpha.
$$
Similarly to (A.6) we obtain
$$
\ {\partial L\over \partial B_\alpha}=
{\rho_{\alpha}{\bfmat{\ell}}{\partial L\over\partial {\bf
V}_{\alpha}}\over
\mu_\alpha}-{{\bfmat{\ell}}{\bfmat{\ell}}^*\over
\mu_\alpha}(\rho_\alpha{\partial L\over\partial {\bf
V}_{\alpha}}{\bf V}_\alpha)+\rho_\alpha R_\alpha
\pmatrix{\displaystyle {{\bfmat{\ell}}{\bfmat{\ell}}^*\over
\mu_\alpha}-B^{-1}_\alpha}.
$$

Consequently,
$$
\det B_\alpha\ { \partial L\over \partial B_\alpha}B_\alpha=
\rho_{0\alpha} \ {\bfmat{\ell} }{\partial L\over\partial {\bf V}_{\alpha}}
B_\alpha-\rho_{0\alpha} \ ({\partial
L\over\partial {\bf V}_{\alpha}} {\bf V}_\alpha )
{\bfmat{\ell} }{\bfmat{\ell} }^*
B_\alpha+\rho_{0\alpha}R_\alpha
({\bfmat{\ell} }{\bfmat{\ell} }^*B_\alpha-\mu_{\alpha} \ I).
$$
If $\lambda_\alpha = t,$
$\displaystyle\ \mu_\alpha=1,B_\alpha=
\left[\matrix{1&0\cr {\bf u}_\alpha&F_\alpha}\right]\ $ and consequently,
$$
T_{0\alpha}=\rho_{0\alpha}\left[\matrix{0&,&-{\bf K}^*_\alpha F_\alpha\cr
\cr 0&,&R_\alpha\ I}\right].
$$
The governing equations of $\alpha$-th component in the Lagrangian
coordinates $\ (t,X_\alpha)\ $ are:
$$
\displaystyle{-\rho_\alpha\theta_\alpha\det B_\alpha\ {\partial
s_{0\alpha}\over \partial t}=0},
$$
$$
\ {\partial \over \partial t}(\rho_{0\alpha}
{\bf K}^*_\alpha F_\alpha)
-\hbox{ div}_\alpha(\rho_{0\alpha}R_\alpha\ I)+R_\alpha\
\displaystyle {\partial \rho_{0\alpha}\over
\partial {\bf X}_\alpha}-\rho_{0\alpha}
\theta_\alpha\ {\partial s_{0\alpha}\over \partial {\bf X}_\alpha}\ \ =\ 0.
$$
Therefore,
$$
{\partial s_{0\alpha}\over \partial t}=0,
\eqno{(D.1)}
$$
$$
{\partial \over \partial t} ({\bf K}^*_\alpha F_\alpha ) -
\nabla_\alpha R_\alpha - \theta_\alpha \nabla_\alpha \ s_{0\alpha}\ =\ 0.
\eqno{(D.2)}
$$
Taking into account the identity
$$
{\partial F_\alpha\over \partial t}={\partial {\bf u}_\alpha\over
{\partial{\bf x}} } F_\alpha
$$
and substituting the partial derivative with respect
to time in the Lagrangian coordinates by the material
derivative in the Eulerian coordinates, we obtain
$$
\displaystyle\ \ {d_\alpha\over dt}s_\alpha =0,\ \ \ \ {\rm with} \ \ \ \
\ \ {d_\alpha\over dt} =
{\partial\over \partial t}\ + \ {\bf u_\alpha}^* {\nabla}^*,
\eqno(D.1')
$$
$$
{d_\alpha\over dt}\
{\bf K}^*_\alpha+ {\bf K}^*_\alpha\ {\partial {\bf u}_\alpha\over \partial
{\bf x}}=\nabla R_\alpha+\theta_\alpha\nabla s_\alpha.
\eqno{(D.2')}
$$
Let us note that (D.1'), (D.2') and the mass conservation laws are
equivalent to (4.5).


\begin{thebibliography}{99}
\bibitem{Gavrilyuk97}
Gavrilyuk, S.L., Gouin, H. and Perepechko, Yu.V., 'Un principe
variationnal pour des m\'elanges de deux fluides',
{\it C.R. Acad. Sci. Paris}
S{\'e}rie II b, {\bf 324} (1997), 483-490.
\bibitem{Gavrilyuk98}
Gavrilyuk, S.L., Gouin, H. and Perepechko, Yu.V., 'Hyperbolic models
of homogeneous two-fluid mixtures', {\it Meccanica}, {\bf 33}
(1998), 161-175.
\bibitem{Serrin59}
Serrin, J., 'Mathematical principles of classical fluid mechanics',
{\it Encyclopedia of Physics}, {\bf VIII/1}, Springer Verlag (1959),
125-263.
\bibitem{Berdichevsky83}
Berdichevsky, V.L., {\it Variational Principles of Continuum
Mechanics}, Moscow, Nauka, 1983.
\bibitem{Geurst86}
Geurst, J.A., 'Variational principles and two-fluid hydrodynamics of
bubbly liquid/gas mixtures', {\it Physica A} {\bf 135} (1986), 455-486.
\bibitem{Geurst85}
Geurst, J.A., 'Virtual mass in two-phase bubbly flow', {\it Physica A}
{\bf 129} (1985), 233-261.
\bibitem{Gouin78}
Gouin, H., 'Contribution \`a une \'etude g\'eom\'etrique et
variationnelle des milieux continus',
{\it Th\`ese d'Etat, Universit\'e de Provence, France}, (1978).
\bibitem{Gouin81}
Gouin, H., 'Lagrangian representation and invariance
properties of perfect fluid flows',
in: {\it Non Linear Problems of Analysis in Geometry and
Mechanics}, Research Notes in Mathematics,
Pitman, {\bf 46} (1981), 128-139.
\bibitem{Serre93}
Serre, D., 'Sur le principe variational des \'equations de la
m\'ecanique des fluides parfaits', {\it Math. Model. Num. Anal.}
{\bf 27}(6) (1993), 739-758.
\bibitem{Gouin87}
Gouin, H., 'Thermodynamic form of the equation of motion for
perfect fluids of grad $n$', {\it C.R. Acad. Sci. Paris}
S{\'e}rie II b, {\bf 305} (1987), 833-838.
\bibitem{Gouin98}
Gouin, H., and Gavrilyuk, S.L., 'Dissipative models of mixtures',
{\it Rendiconti del circolo matematico di Palermo}, serie  II, Suppl. \textbf{78} (2006), 133-145.
\bibitem{Stewart84}
Stewart, H.B., and Wendroff, B., 'Two-phase flow: models and
methods', {\it Journal of Comput. Physics }
{\bf 56} (1984), 363-409.
\bibitem{Prosperetti59}
Prosperetti, A., and Satrape, J.V., 'Stability of
two-phase flow models', in: Joseph, D.D. \& Schaeffer, D.G., (Ed.)
{\it Two-Phase Flows and Waves}, Springer Verlag (1990), 98-117.
\bibitem{Gouin90}
Gouin, H., 'Variational theory of mixtures in continuum mechanics',
{\it Eur. J. Mech}, B/Fluids {\bf 9} (1990), 469-491.
\end{thebibliography}
\end{document}